\documentclass{amsart}

\usepackage{amsmath}                        
\usepackage{amssymb}                        
\usepackage{dsfont} 	                    
\usepackage{bm}                             
\usepackage{mathtools}                      
\usepackage[hypertexnames=false]{hyperref}  
\usepackage[scientific-notation=true]{siunitx}
\usepackage{blkarray}
\usepackage{enumerate}
\usepackage{centernot}
\usepackage{paralist}
\usepackage{stackrel}
\usepackage{multirow}
\usepackage[table,xcdraw]{xcolor}
\usepackage{booktabs}
\usepackage[colorinlistoftodos,prependcaption,textsize=tiny]{todonotes} 
\usepackage{pagenote}
\usepackage{setspace}
\usepackage[english]{babel}

\makepagenote

\renewcommand*{\pagenotesubhead}[2]{}
\let\footnote\pagenote

\usepackage{algorithm}
\usepackage{algpseudocode}

\usepackage{listings}
\usepackage{color}

\usepackage{amsthm}
\theoremstyle{plain}
\newtheorem{definition}{Definition}
\theoremstyle{definition}
\newtheorem{example}{Example}

\theoremstyle{remark}

\newcommand*{\QEDE}{\hfill\ensuremath{\blacksquare}}

\makeatletter
\newcommand{\addresseshere}{%
  \enddoc@text\let\enddoc@text\relax
}

\definecolor{lightgrey}{rgb}{0.9,0.9,0.9}
\definecolor{darkgreen}{rgb}{0,0.6,0}
\calclayout

\begin{document}

\title[Social Network Analysis of Corruption]
{Social Network Analysis of Corruption Structures: Adjacency Matrices Supporting the Visualization and Quantification of Layeredness}

\author[C.F.W.\ Peeters]{Carel F.W.\ Peeters}
\address[Carel F.W.\ Peeters]{
Dept.\ of Epidemiology \& Biostatistics \\
Amsterdam Public Health research institute \\
Amsterdam University medical centers, Location VUmc \\
Amsterdam\\
The Netherlands}
\email{cf.peeters@amsterdamumc.nl}

\maketitle


\section{Introduction}
\label{SEC:Intro}
Corruption and other integrity violations have been central themes in the body of work produced by Leo Huberts.
From early reports \cite{Leo98,Leo2000} to some of the latest publications \cite{Leo18,GHS18} they have dominated his thinking.
In these works integrity is seen as an encompassing quality, one that should compel any public official to act ``in accordance with the [culturally] relevant moral values, norms and rules" \cite{HLPmeas}.
Any act in discordance would then be an integrity violation.
One of the most important contributions of Leo lies in offering a typology of (categories of) integrity violations \cite{LPS99}.
This typology represents a slippery slope ranging from private-life misconduct to corruption.
It conveys that public misconduct has a clear moral dimension and is a much more diverse phenomenon than corruption alone.

Nevertheless, as corruption is the pinnacle of public misconduct, it tends to catch the attention of both the media and the public.
Here, we will indeed focus on corruption, in particular as defined in many penal codes.
That is, the misuse of (public) power for private gain, especially through involvement in bribing.
Often, corruption is described as taking place within or supported by a network \cite{DV99,Hill10,H14,Osifo18,Slinger18}:
A collection of individuals structured in such a way as to enable the transaction of bribes for favors.
Surprisingly, despite the network nomenclature, corruption is rarely analyzed from the network perspective using the tools of network science.
Notable exceptions are Lauchs \emph{et al.}\ \cite{LKY11}, Chang \cite{C18}, Divi\'{a}k \emph{et al.}\ \cite{DDS18}, and Ribeiro \emph{et al.}\ \cite{RibETAL18}.
This is a missed opportunity, as leveraging the tools of networks science enables one to exploit a mechanistic perspective on corruption and its social dynamics and may ultimately lead to deeper insight into its structural aspects.

Here, we will argue that analyzing corruption from the perspective of network science is beneficial to its understanding.
Section \ref{SEC:Networkprimer} gives a very short introduction to a selection of core concepts from network science.
It especially focuses on the evaluation of the structural importance of actors within the network and multiplex networks, i.e., layered networks in which ties may signify multiple relationship types.
Section \ref{SEC:Rath} gives the background to the affair used to exemplify network analyzes: the Czech Rath affair.
It makes clear that a mechanistic understanding of corruption hinges upon a multi-relational approach to networks and their analysis.
Section \ref{SEC:SDAM} proposes a simple approach to evaluate and visualize actor-importance in multi-relational networks.
This approach is then used to analyze the corruption-structure of the Rath affair in Section \ref{SEC:AAffair}.
Section \ref{SEC:DISCUSS} closes with a discussion.

\section{A (very short) primer on network analysis}
\label{SEC:Networkprimer}
A network is a collection of nodes (or vertices) connected to each other in a certain way through a collection of edges (or ties).
As we are interested in corruption, the nodes represent actors or persons, and the edges represent relational ties of a certain type.
Such networks are usually termed \emph{social networks}.
The structure of the connections among the nodes is called the \emph{topology} of the network, and it contains much information on the mechanisms of action and the flow of information.
While (social) network analysis is a large and growing field, we will concern ourselves with certain specific concepts of interest to the current work: adjacency matrices for undirected networks, measures of centrality, and multiplexity.

A network is thus a graphical object $\mathcal{G} = (\mathcal{V}, \mathcal{E})$ consisting of a finite set $\mathcal{V} = \{Y_{1},\ldots,Y_{p}\}$ of nodes and set of edges $\mathcal{E}$.
Edges in $\mathcal{E}$ consist of pairs of nodes that are connected.
Mathematically, a network can be encoded in an adjacency matrix $\mathbf{A}$.
These are square matrices, i.e., matrices in which the number of rows equals the number of columns.
The rows and columns represent the nodes and the entries $(\mathbf{A})_{ij}$ represent the status of the connections from node $Y_i$ to node $Y_j$.
We are interested in unweighted, undirected graphs in which self-loops are not considered.
This means that $(\mathbf{A})_{ii} = 0$, and $(\mathbf{A})_{ij} = (\mathbf{A})_{ji}$ according to:
\begin{equation}\label{Eq: DiffAdjExp1}
    (\mathbf{A})_{ij} = \begin{cases}
                          1 & \mbox{if} ~Y_{i} - Y_{j}\\
                          0 & \mbox{if} ~Y_{i} \centernot{-} Y_{j}
                        \end{cases}.
\end{equation}
Hence, $(\mathbf{A})_{ij} = (\mathbf{A})_{ji} = 1$ if there is an undirected edge between nodes $Y_{i}$ and $Y_{j}$, i.e., $Y_{i} - Y_{j}$.
And $(\mathbf{A})_{ij} = (\mathbf{A})_{ji} = 0$ if there is no edge between nodes $Y_{i}$ and $Y_{j}$, i.e., $Y_{i} \centernot{-} Y_{j}$.
The adjacency matrix thus holds the information for the construction of the network topology and it can be used to quantify network characteristics at the node, path (a sequence of connected nodes that, when traveled, never traverses a node twice), community (a collection of closely connected nodes), and global level.

We will focus mostly on the quantification of node characteristics.
A key concept is \emph{centrality}, referring to the (structural) importance of a node.
Understanding networks and network mechanisms at the node level thus usually revolves around quantifying and ranking each node's centrality.
Centrality may be quantified in many ways.
We will consider the simplest and most used quantification, \emph{degree}:
\begin{equation}\label{Eq:Central}
    c_i = \sum_j ({\mathbf{A}})_{ij}.
\end{equation}
Hence, the degree centrality of node $i$ ($c_i$) is simply the number of ties associated with that node and can be obtained by summing over the column elements in the appropriate row of the adjacency matrix defined in (\ref{Eq: DiffAdjExp1}).
The degree expresses how connected a node is within the network and actors with high degree are thought of as being (structurally) important.

The ties in a single network are usually used to encode a certain well-defined type of relationship, such as `is acquainted with' or `is a colleague of'.
A defining aspect of corruption (and many other) structures is that there are multiple types of relationship within the network that are important to its understanding.
In that case we have the same nodes, but the network is layered with each layer representing a relationship-type.
The network topology is then dependent on the layer.
Such networks are known as \emph{multiplex}, \emph{multi-relational}, or \emph{layered} networks.
We will use these terms interchangeably.
The quantification of characteristics such as centrality is more problematic in layered networks \cite{Bocca14,RankingNC}.
The simplest approach would be to calculate a centrality measure for each node within each layer separately, followed by a summarization of sorts over the layers such as their correlation \cite{Bocca14}.
Such approaches may fail to capture structural changes in node-centrality when moving from one layer to the next \cite{NL15}.
In Section \ref{SEC:SDAM} a simple quantification of pairwise structural degree change and stability is introduced.
This approach to quantification will be used in the analysis (Section \ref{SEC:AAffair}) of the multiplex corruption network introduced in Section \ref{SEC:Rath}.

Example \ref{OverviewExample} exemplifies the concepts touched upon above with simple toy-networks.
For an overview of other concepts in social network analysis and network science in general we confine by referring to Borgatti \emph{et al.} \cite{BorgattiASN} and Newman \cite{NewmanBook}.

\begin{example}
\label{OverviewExample}
\begin{sloppypar}
Say we have two graphical objects, $\mathcal{G}_1$ and $\mathcal{G}_2$, where $\mathcal{G}_1 = (\mathcal{V}, \mathcal{E}_1)$ and $\mathcal{G}_2 = (\mathcal{V}, \mathcal{E}_2)$.
Hence, the graphs have the same node-set, but different edge-sets.
We thus have a multiplex network.
Let $\mathcal{V} = \{\mathrm{A, B, C, D, E, F}\}$.
We thus have 6 actors indicated by capital letters.
Also, let the edge-set for $\mathcal{G}_1$ be given as $\mathcal{E}_1 = \{\mathrm{(A,B), (A,C), (A,D), (A,E), (A,F), (B,C), (B,F), (C,D), (D,E), (E,F)}\}$ while the edge-set for $\mathcal{G}_2$ is $\mathcal{E}_2 = \{\mathrm{(B,C), (B,F), (C,D), (D,E), (E,F)}\}$.
These graphical objects can be encoded in the following corresponding relation-specific adjacency matrices:
\begin{equation}\nonumber
\begin{array}{c c} &
\begin{array}{c c c c c c} A & B & C & D & E & F \\
\end{array}
\\
\mathbf{A}_1 =
\begin{array}{c c}
A\\
B\\
C\\
D\\
E\\
F
\end{array}
&
\left[
\begin{array}{c c c c c c}
0 & 1 & 1 & 1 & 1 & 1 \\
1 & 0 & 1 & 0 & 0 & 1 \\
1 & 1 & 0 & 1 & 0 & 0 \\
1 & 0 & 1 & 0 & 1 & 0 \\
1 & 0 & 0 & 1 & 0 & 1 \\
1 & 1 & 0 & 0 & 1 & 0 \\
\end{array}
\right]
\end{array},
\begin{array}{c c} &
\begin{array}{c c c c c c} A & B & C & D & E & F \\
\end{array}
\\
\mathbf{A}_2 =
\begin{array}{c c}
A\\
B\\
C\\
D\\
E\\
F
\end{array}
&
\left[
\begin{array}{c c c c c c}
0 & 0 & 0 & 0 & 0 & 0 \\
0 & 0 & 1 & 0 & 0 & 1 \\
0 & 1 & 0 & 1 & 0 & 0 \\
0 & 0 & 1 & 0 & 1 & 0 \\
0 & 0 & 0 & 1 & 0 & 1 \\
0 & 1 & 0 & 0 & 1 & 0 \\
\end{array}
\right]
\end{array}.
\end{equation}
The accompanying networks are visualized in Figure \ref{FIG:ToyNetworks}.
\begin{figure}[h!]
\centering
  \includegraphics[width=.85\textwidth]{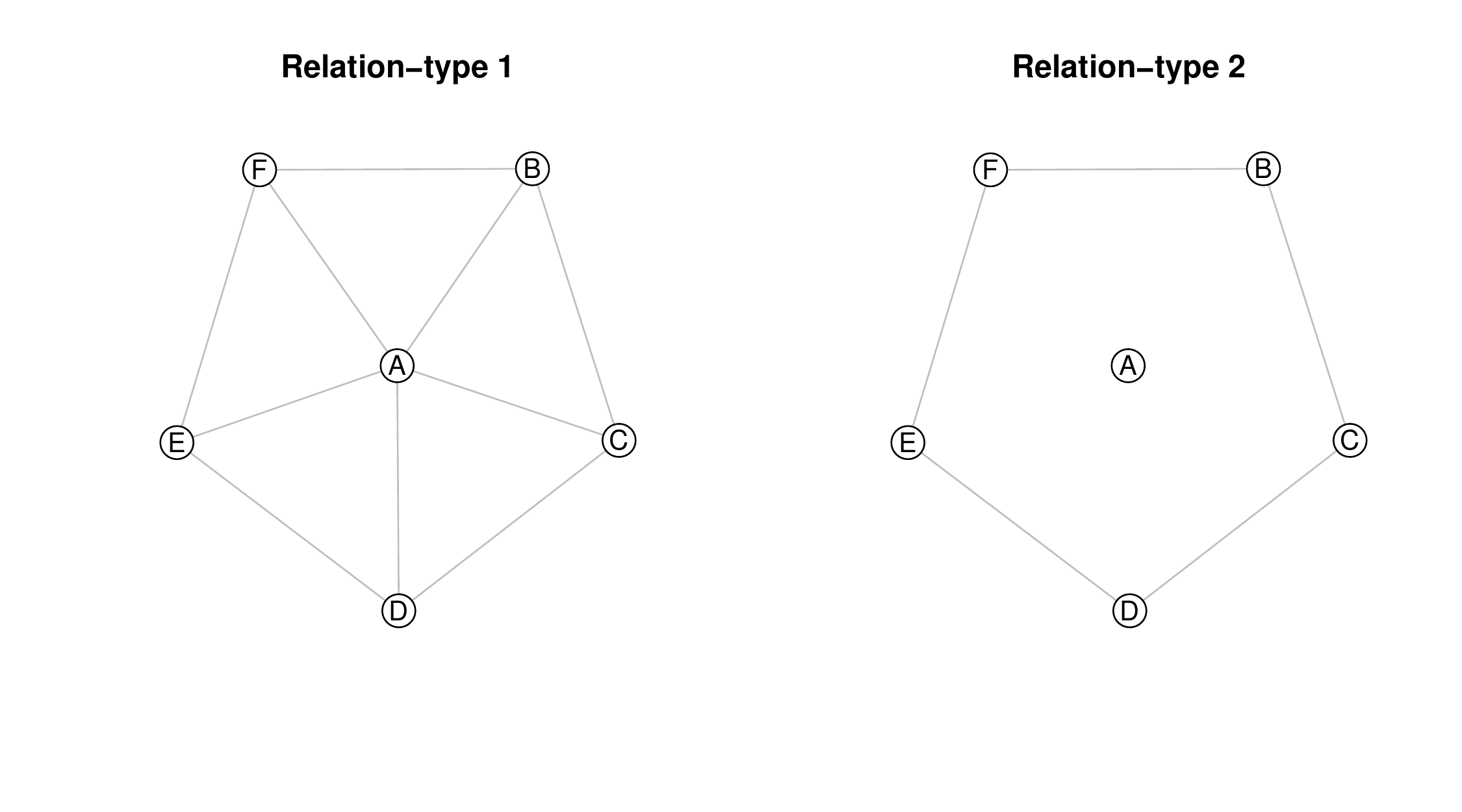}
    \caption{\footnotesize{Visualized relation-specific networks stemming from Example \ref{OverviewExample}.}}
  \label{FIG:ToyNetworks}
\end{figure}

For example, we see that the unordered pair $\mathrm{(A,C)}$ is an element in edge-set $\mathcal{E}_1$.
Hence, $({\mathbf{A}}_1)_{\mathrm{AC}} = ({\mathbf{A}}_1)_{\mathrm{CA}} = 1$, i.e., the element in the A-row and C-column as well as the element in the C-row and A-column of the adjacency matrix for relation-type 1 $({\mathbf{A}}_1)$ is 1.
In the corresponding network visualization we indeed see an edge between nodes A and C.
The unordered pair $\mathrm{(A,C)}$ is \emph{not} an element in edge-set $\mathcal{E}_2$ and, hence, $({\mathbf{A}}_2)_{\mathrm{AC}} = ({\mathbf{A}}_2)_{\mathrm{CA}} = 0$.
There is also no edge between nodes A and C in the network for relation-type 2.
We see that node A has the highest degree centrality in the first network: 5, as A has 5 connections to other nodes.
Node A has the lowest degree of 0 in the second network as it is unconnected for relation-type 2.
Our actors could be certain employees in a large firm, and relation-type 1 could signify whether there is `contact at work', while relation-type 2 signifies `contact outside of work'.
Actor A, apparently, is in contact with many people at work but not so much outside of the work-environment.
The other actors all see the other persons they know at work outside of the work-environment, except for actor A (how sad for A).
\QEDE
\end{sloppypar}
\end{example}

\section{The Rath affair}
\label{SEC:Rath}

\subsection{Background}
\label{SSEC:RathBackground}
On 14 May 2012, David Rath, Governor of the Central Bohemian Region in the Czech Republic, Member of the Czech parliament, and former minister of Health, was caught red-handed carrying a wine-box stuffed with bribe money.
This arrest came at the hands of an intensive police operation.
Next to Rath, Katerina Pancova and Petr Kott also were arrested.
In addition to being a couple, Pancova and Kott were also close (political) friends and colleagues of Rath.
They were accused of manipulating public contracts and accepting bribes, especially in relation to the renovation of public estates and the acquisition of medical equipment for hospitals.
Many of the manipulated contracts involved projects that were to be partly financed with EU-funds.
In the weeks that followed eight more people were arrested in the ongoing investigation.
These people were mainly representatives and managers of construction and healthcare companies that were conducive in the manipulation of public contracts.
The affair turned out to be one of the most publicized corruption scandals in Czech history and, owing to Rath being a household name, came to be known as the Rath affair.
All those arrested were eventually sentenced to (probationary) incarceration and/or monetary fines.
Table \ref{TABLE:Actors} contains an overview of the actors involved and their roles in the timeline preceding the arrests.
For more information about the background of the Rath affair, see \cite{DDS18,PeopleOverviewGolis}.

\begin{table}[]
\caption{Overview of the actors and their roles in the timeline preceding the arrests.}
\label{TABLE:Actors}
\resizebox{\textwidth}{!}{%
\begin{tabular}{lll}
\toprule
Actor                                                      &  & Role in 2012                                                                                       \\ \midrule
                                                           &  & Medical doctor by training                                                                         \\
                                                           &  & Deputy of the Social Democratic Party (CSSD) and former member of the Civic Democratic Party (ODS) \\
                                                           &  & Member of the Czech parliament                                                                     \\
\multirow{-4}{*}{David Rath}                               &  & Governor of the Central Bohemian Region                                                            \\
\rowcolor[HTML]{C0C0C0}
\cellcolor[HTML]{C0C0C0}                                   &  & Medical doctor by training and former colleague of Rath                                            \\
\rowcolor[HTML]{C0C0C0}
\cellcolor[HTML]{C0C0C0}                                   &  & Director of the Kladno hospital                                                                    \\
\rowcolor[HTML]{C0C0C0}
\cellcolor[HTML]{C0C0C0}                                   &  & CSSD regional committee member                                                                     \\
\rowcolor[HTML]{C0C0C0}
\multirow{-4}{*}{\cellcolor[HTML]{C0C0C0}Katerina Pancova} &  & Wife of Petr Kott                                                                                  \\
                                                           &  & Medical doctor by training                                                                         \\
                                                           &  & Politician, formerly associated with the ODS                                                       \\
                                                           &  & Employee of companies owned by David Rath                                                          \\
\multirow{-4}{*}{Petr Kott}                                &  & Husband to Katerina Pancova                                                                        \\
\rowcolor[HTML]{C0C0C0}
Lucia Novanska                                             &  & Manager government contracts at Trust Office ML Compet                                             \\
Ivana Salacova                                             &  & Representative to construction company Fisa                                                        \\
\rowcolor[HTML]{C0C0C0}
Pavel Drazdansky                                           &  & Manager of construction company Konstruktiva Branko                                                \\
Tomas Mlady                                                &  & Manager of construction company Konstruktiva Branko                                                \\
\rowcolor[HTML]{C0C0C0}
Vaclav Kovanda                                             &  & Manager of construction company Pohl                                                               \\
Martin Jires                                               &  & Manager medical (equipment) company Puro-Klima                                                     \\
\rowcolor[HTML]{C0C0C0}
Jindrich Rehak                                             &  & Executive director Hospimed Healthcare Company                                                     \\
Jan Hajek                                                  &  & Manager medical company B. Braun Medical                                                           \\ \bottomrule
\multicolumn{3}{l}{}\\[-0.75\normalbaselineskip]%
\multicolumn{3}{l}{Source: \cite{PeopleOverviewGolis}.}%
\end{tabular}%
}
\end{table}

\subsection{Data}
\label{SSEC:RathData}
The data was collected by Divi\'{a}k \emph{et al.} \cite{DDS18}.
The actors involved in their network-data collection were those who were eventually charged in connection with the Rath affair and span the people mentioned in Table \ref{TABLE:Actors}.
They used publicly available open-source online and print media as their data source \cite{Newton}.
After identifying relevant articles based on strategic search terms, ties between any pair of involved actors were recorded when an article contained a statement on their connection.
Corruption networks are inherently multi-relational and ties for three relationship-types were coded \cite{DDS18}:
\begin{enumerate}
  \item Mutual involvement preceding the affair, such as shared previous affiliations: \emph{pre-existing relations};
  \item Mutual involvement in the transfer of (monetary) resources: \emph{resource transfer relations};
  \item Mutual communication or cooperation not involving direct resource transfer: \emph{collaboration relations}.
\end{enumerate}
All relationship-types are, for present purposes, considered to be undirected.
For more information regarding data collection, search terms and eligibility criteria, see Divi\'{a}k \emph{et al.} \cite{DDS18}.

In the next section we will develop a very simple framework for the visualization and analysis of (node-importance in) multilayered networks.
In Section \ref{SEC:AAffair} we will use this framework to visualize and analyze the Rath affair data.
These visualizations and analyzes may be considered to be additions to the excellent analysis performed by Divi\'{a}k \emph{et al.} \cite{DDS18}.

\section{Shared and differential adjacency matrices}
\label{SEC:SDAM}
\subsection{Basic idea}
Our goal is insightful visualization of multiplexity and simple quantification of centrality in multiplex structures.
We will do so by visualizing and quantifying shared and differential ties, between any pairing of relationship-types.
This requires some simple extensions to the basic adjacency matrix.

For visualizing and quantifying differential ties we define the (pairwise) \emph{differential adjacency matrix}.
A differential adjacency matrix between any two relationship-types is simply the matrix subtraction between the corresponding relation-specific adjacency matrices.
This will result in a signed adjacency matrix, in which an entry is $1$ if a tie is unique to one relationship-type but not the other and $-1$ when vice versa.
An entry is then $0$ if a tie is either present or absent in both relationship-types.
For visualizing and quantifying shared ties we define the (pairwise) \emph{shared adjacency matrix}.
A shared adjacency matrix between any two relationship-types has entries of $1$ when a tie is shared between the relations and $0$ otherwise.
See Appendix \ref{APP:AdjacencyMatrices} for more formal definitions of the differential and shared adjacency matrix.

\subsection{Visualization}
The shared and differential adjacency matrices can be used to visualize shared and differential networks over pairs of relation-specific graphs.
The differential adjacency matrix encodes for the differential network between pairs of relation-types.
The sign of the entries is now an attribute of the corresponding ties.
Hence, this information can be added to the graph by, for example, edge-colorings.
The coloring is then indicative of the relationship-type to which the tie is unique.
In a sense, the differential graph depicts the `differential wiring' between relationship-types.
To juxtapose the differential wiring with the shared wiring it is also insightful to visualize the shared network.
This is simply the network corresponding to the shared adjacency matrix.
For maximum visual comparability it is recommended to visualize the shared and differential networks with the same node-coordinates.
Note that, when node-attributes are available, these can also be used for node-coloring to embed additional information in the visualization.

\subsection{Centrality quantification}
We also desire, next to having a simple visual representation of shared and differential networks over relationship-pairings, to quantify shared and differential centrality.
One way to do this would be to calculate the multiplex participation coefficient \cite{Battiston}.
This measure classifies each node, according to the level of uniformity of its connectedness over layers, into one of 3 classes.
These classes indicate increasing levels of uniformity: focused, mixed, and multiplex.
While this measure can describe the dispersion of connections over the layers, it does not capture the depth of shared and differential connections across layers.
For example, if we have 3 layers and a specific node has 3 connections in all 3 layers, then this node would attain the maximum participation coefficient and would be considered multiplex.
For a mechanistic understanding of this node across layers, it is however important to understand whether its connections change or behave stably across layers.
Moreover, if in a well-connected network a node would have just one tie in all layers it would also attain the maximum participation coefficient, even though it would be a marginal node (in terms of degree) across the layers.
Hence, we seek a measure to add additional information to quantifications such as the relation-specific degrees and the multiplex participation coefficient.

We especially seek to quantify stable and differential connectedness across pairs of layers.
For this we can use the defined adjacency matrices.
Let $\tilde{\mathbf{A}}$ be a generic indication of either the differential or shared adjacency matrix (Appendix \ref{APP:AdjacencyMatrices}).
To evaluate shared and differential degree centrality one would then replace $\mathbf{A}$ by the absolute representation of $\tilde{\mathbf{A}}$ in (\ref{Eq:Central}):
\begin{equation}\label{Eq:Central2}\nonumber
    \tilde{c}_i = \sum_j |(\tilde{\mathbf{A}})_{ij}|.
\end{equation}
Note that we are summing over the absolute entries in order to avoid the sign attribute in the differential adjacency matrix to nullify the degree measure.
For any pair of layers we may then use the shared and differential degree centralities to (qualitatively) assess whether a node behaves between these layers as one of four archetypes.
The node could be loosely connected (both the shared and differential degrees are relatively low), stably connected (shared degree relatively high and differential degree relatively low), differentially connected (shared degree relatively low and differential degree relatively high), or both stably and differentially connected (both the shared and the differential degree relatively high).
Before turning to the analysis of the Rath affair we provide an additional example to exemplify these new concepts.

\begin{example}
\label{OverviewExampleExtended}
\begin{sloppypar}
Consider the relation-specific networks from Example \ref{OverviewExample}.
Using the principles laid out previously, we can obtain the corresponding differential and shared network topologies between relationship-types 1 and 2.
These are visualized in Figure \ref{FIG:ToyNetworksShareDiff}.
\begin{figure}[h!]
\centering
  \includegraphics[width=.85\textwidth]{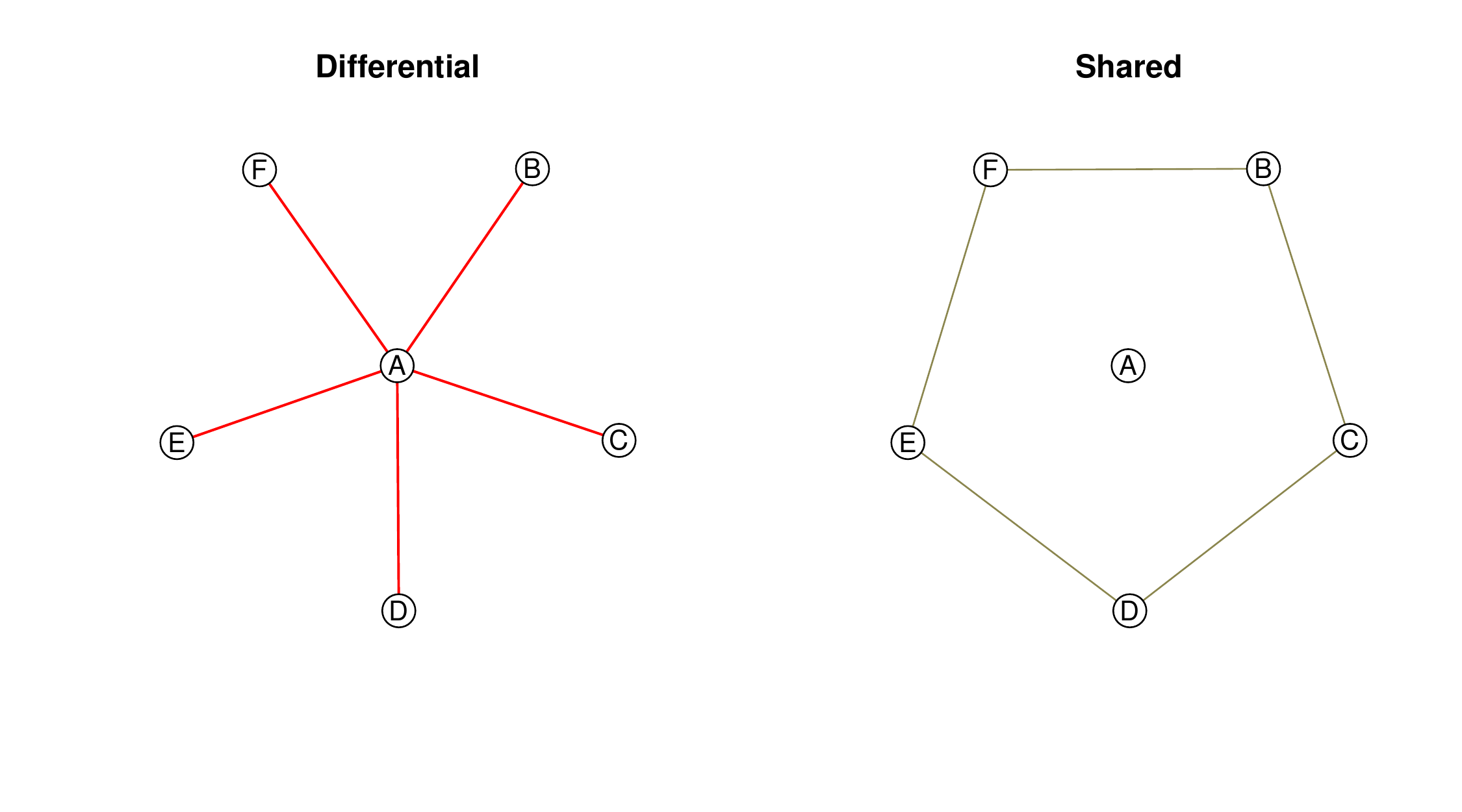}
    \caption{\footnotesize{
    \emph{Left-panel}: Differential network between relationship-types 1 and 2.
    \emph{Right-panel}: Shared network between relationship-types 1 and 2.}}
  \label{FIG:ToyNetworksShareDiff}
\end{figure}
The left-hand panel contains the differential network between relationship-types 1 and 2.
Red ties are unique to relationship-type 1.
We immediately see that there are no ties unique to relationship-type 2.
The right-hand panel then contains the shared network.
We immediately see that all actors except A are stably connected over the two layers.
Hence, they can be said to behave in a multiplex manner.
From a differential roles perspective though, actor A would be the most interesting actor.
It is relationally the most differential actor.
These observations are directly reflected in the corresponding shared and differential degree centralities.
\QEDE
\end{sloppypar}
\end{example}

\section{Analysis of the Rath affair}
\label{SEC:AAffair}
\subsection{Relation-specific layers}
\begin{figure}[h!]
\centering
  \includegraphics[width=\textwidth]{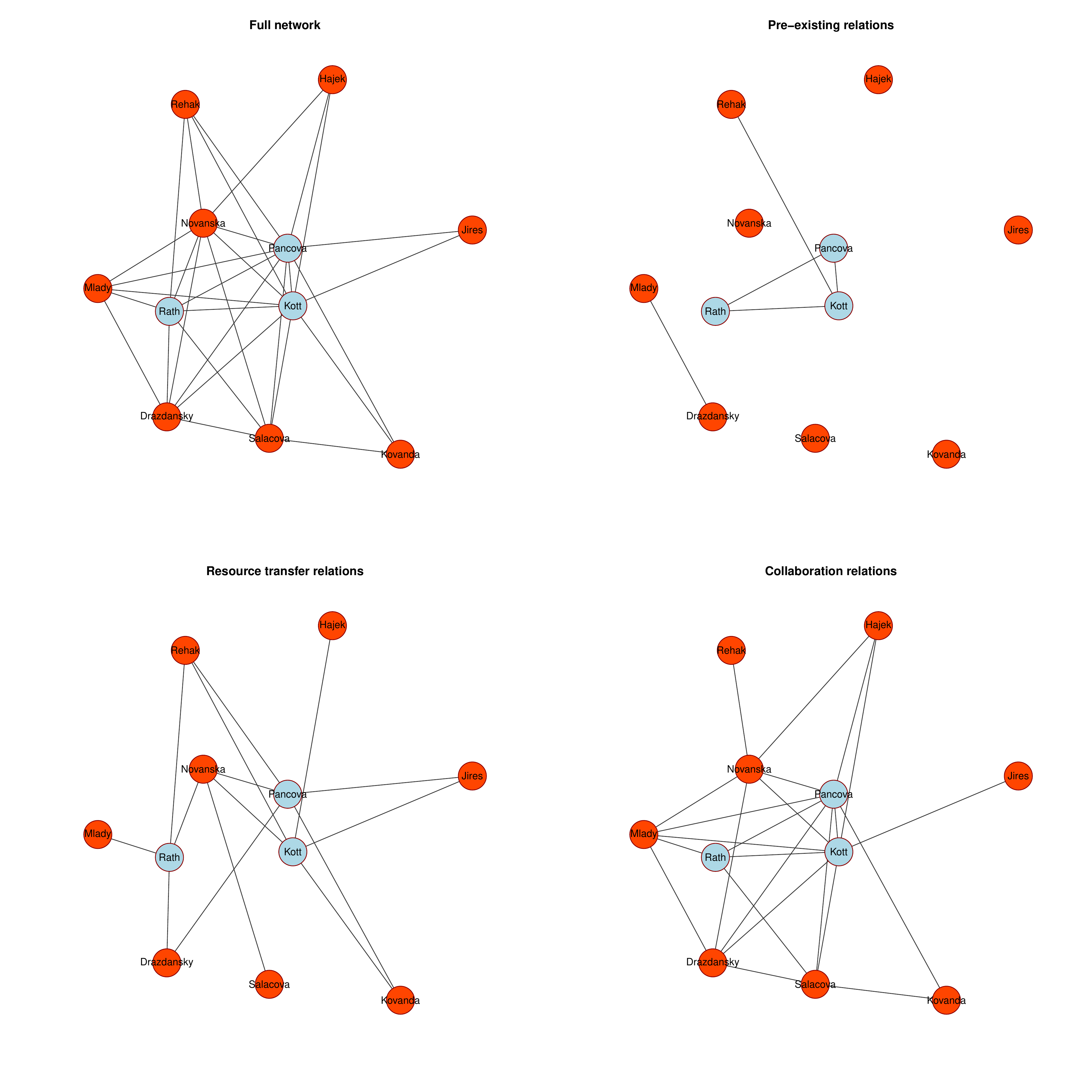}
    \caption{\footnotesize{Full and relation-specific network topologies.
    \emph{Top-left panel}: full network.
    \emph{Top-right panel}: pre-existing relations network.
    \emph{Bottom-left panel}: resource transfer relations network.
    \emph{Bottom-right panel}: collaboration relations network.
    The blue nodes represent the political actors while the red nodes represent the business actors.}}
  \label{FIG:RathNetworks}
\end{figure}

We will start our analysis by assessing the full network-structure as well as the relation-specific layers.
We will focus on the simple network-concepts introduced so far.
For an extensive analysis with alternative measures see Divi\'{a}k \emph{et al.} \cite{DDS18}.
Figure \ref{FIG:RathNetworks} contains a visualization of both the full network and the relation-specific networks.

The ties in the full network are based on presence in the relation-specific networks.
Hence, if a tie is present in at least one of the relation-specific networks it is present in the full network.
The full network was visualized with the Fruchterman-Reingold algorithm \cite{FRalgo}, which positions the nodes such that all ties are approximately of equal length while minimizing the number of crossing ties.
It also prefers coiled structures, tending to place highly connected nodes towards the center of the network.
The node-coordinates of the full network then serve as the reference coordinates for the relation-specific networks.
The nodes are labeled with the names of the actors.
Moreover, the nodes are colored according to the actor-type attribute: the blue nodes represent the political actors while the red nodes represent the business actors.
Corresponding degree centralities for each of the networks can be found in the second to fifth columns of Table \ref{TABLE:Degrees}.

The full network is quite densely connected.
This is mainly due to the density of connections in the resource transfer and collaboration layers.
The pre-existing relations network is less densely connected, with pre-existing ties limited mostly to the political actors.
From a degree centrality perspective, the political actors are central in the pre-existing layer, Lucia Novanska is additionally important in the resource transfer layer, while especially Novanska, Pancova and Kott are central in the collaborations layer.
The centrality of Novanska in the resource transfer and collaboration layers is natural as her role as manager of government contracts at a trust office made her a broker of sorts between many of the business actors and the political actors.

We now make several additional observations not directly made by Divi\'{a}k \emph{et al.} \cite{DDS18}.
First, the full network operates as a single well-connected module from the community-finding perspective (results not shown, but can be obtained from the coding script; see Appendix \ref{APP:SoftData}).
That is, when trying to divide the full network into collections of nodes \cite{GNalgo} that are naturally grouped (according to the density of their ties) we cannot subdivide beyond the full network.
This hinges mainly on the structure of the collaboration layer, which also operates as a single well-connected module.
The pre-existing and resource transfer layers can be subdivided into several modules.
Second, from the perspective of the node attribute (political or business actor) the resource transfer layer has a near-bipartite structure.
That is, the ties are predominantly between business and political actors, not among political or business actors.
This is a natural and possibly endemic characteristic of corruption structures: as political and business actors are seeking mutual benefits, resource transfers occur mostly between these actor sets rather than within.

\subsection{Shared and differential layeredness}
\begin{figure}[h!]
\centering
  \includegraphics[width=.92\textwidth]{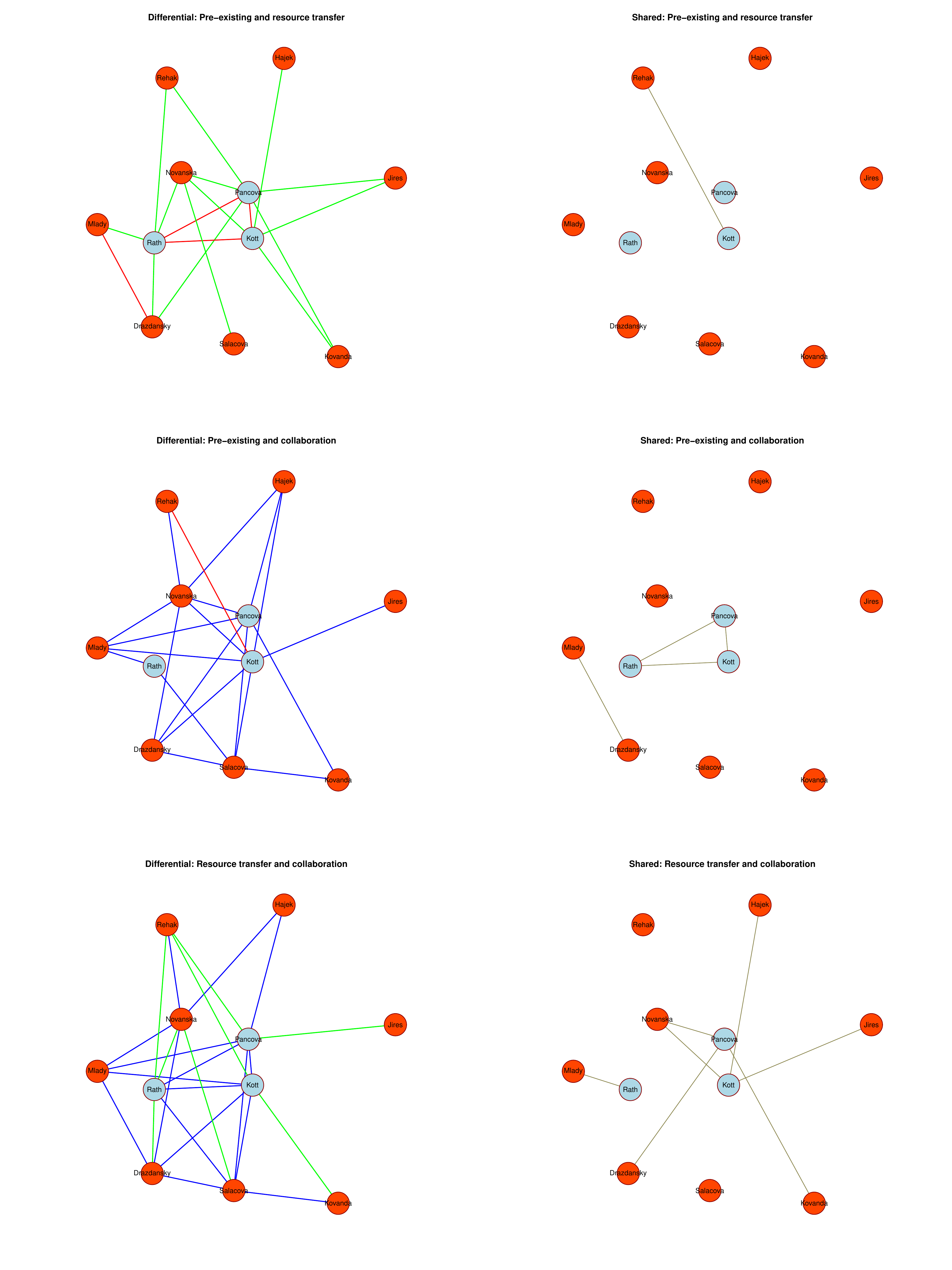}
    \caption{\footnotesize{Shared and differential networks between pairs of relationship-types.
    \emph{Top panels}: Differential (left) and shared (right) networks between the pre-existing and resource transfer relationship-types.
    \emph{Middle panels}: Differential (left) and shared (right) networks between the pre-existing and collaboration relationship-types.
    \emph{Bottom panels}: Differential (left) and shared (right) networks between the resource transfer and collaboration relationship-types.
    Again, the blue nodes represent the political actors while the red nodes represent the business actors.}}
  \label{FIG:RathNetworksDiffShare}
\end{figure}

Figure \ref{FIG:RathNetworksDiffShare} contains a visualization of the shared and differential networks between pairs of relationship-types.
Again, the visualizations use the node-coordinates of the full network.
The corresponding shared and differential degree centralities can be found in the sixth to eleventh columns of Table \ref{TABLE:Degrees}.

The top panels of Figure \ref{FIG:RathNetworksDiffShare} represent the differential (left) and shared (right) networks between the pre-existing and resource transfer relationship-types.
Red ties are unique to the pre-existing relations network while green ties are unique to the resource transfer relations network.
Khaki ties (right panel) indicate the ties shared between the pre-existing and resource transfer relationship-types.
All nodes between these layers are predominantly differentially connected nodes.
This is mainly due to the very limited overlap in connections between the layers.
The most differentially connected actors are the three political actors.
They each have multiple resource transfer ties to business actors, which again highlights the (near) bipartite structure of these relations.

The middle panels of Figure \ref{FIG:RathNetworksDiffShare} represent the differential (left) and shared (right) networks between the pre-existing and collaboration relationship-types.
Red ties are unique to the pre-existing relations network while blue ties are unique to the collaboration relations network.
Khaki ties (right panel) indicate the ties shared between the pre-existing and collaboration relationship-types.
The triangular connectedness of the political actors is a (stable) shared motif between these layers.
The most differentially connected actors are Pancova, Kott, and Novanska.

The bottom panels of Figure \ref{FIG:RathNetworksDiffShare} represent the differential (left) and shared (right) networks between the resource transfer and collaboration relationship-types.
Green ties are unique to the resource transfer relations network while blue ties are unique to the collaboration relations network.
Khaki ties (right panel) indicate the ties shared between the resource transfer and collaboration relationship-types.
The most stably connected actors are Pancova, Kott, and Novanska.
The most differentially connected actors are the former joined by Rath and Salacova.

From a qualitative standpoint one could argue that those actors that are both stably and differentially connected across layers are the most important actors.
Their stable connections provide binding elements between layers while their differential connections provide relation-specific expansions that add to the density of the full network operation.
From this perspective especially Pancova and Kott, and to a somewhat lesser degree Novanska and Rath, are the core actors.
It has been noted previously that Pancova and Kott took care of much of the work within the network \cite{DDS18} while Novanska earned her centrality as a broker of sorts.
Drazdansky and Salacova are then semi-core actors, being well-connected mainly due to their many collaboration ties.
The remaining actors may then be considered more peripheral.
In the next subsection we will cast our qualitative observations into some conjectures.

\begin{table}[]
\caption{Full, relation-specific, differential, and shared degree centralities.}
\label{TABLE:Degrees}
\resizebox{\textwidth}{!}{%
\begin{tabular}{llrrrrlrrrlrrr}
\toprule
           &  & \multicolumn{1}{l}{}     & \multicolumn{1}{l}{}    & \multicolumn{1}{l}{}   & \multicolumn{1}{l}{}   &  & \multicolumn{3}{c}{Differential}                                                             &  & \multicolumn{3}{c}{Shared}                                                                   \\ \cline{8-10} \cline{12-14}
           &  & \multicolumn{1}{l}{Full} & \multicolumn{1}{l}{Pre} & \multicolumn{1}{l}{RT} & \multicolumn{1}{l}{Co} &  & \multicolumn{1}{c}{Pre -- RT} & \multicolumn{1}{c}{Pre -- Co} & \multicolumn{1}{c}{RT -- Co} &  & \multicolumn{1}{c}{Pre -- RT} & \multicolumn{1}{c}{Pre -- Co} & \multicolumn{1}{c}{RT -- Co} \\ \midrule
Pancova    &  & 10                       & 2                       & 5                      & 8                      &  & 7                             & 6                             & 7                            &  & 0                             & 2                             & 3                            \\
Kott       &  & 10                       & 3                       & 5                      & 8                      &  & 6                             & 7                             & 7                            &  & 1                             & 2                             & 3                            \\
Novanska   &  & 8                        & 0                       & 4                      & 6                      &  & 4                             & 6                             & 6                            &  & 0                             & 0                             & 2                            \\
Rath       &  & 7                        & 2                       & 4                      & 4                      &  & 6                             & 2                             & 6                            &  & 0                             & 2                             & 1                            \\
Drazdansky &  & 6                        & 1                       & 2                      & 5                      &  & 3                             & 4                             & 5                            &  & 0                             & 1                             & 1                            \\
Salacova   &  & 6                        & 0                       & 1                      & 5                      &  & 1                             & 5                             & 6                            &  & 0                             & 0                             & 0                            \\
Mlady      &  & 5                        & 1                       & 1                      & 5                      &  & 2                             & 4                             & 4                            &  & 0                             & 1                             & 1                            \\
Rehak      &  & 4                        & 1                       & 3                      & 1                      &  & 2                             & 2                             & 4                            &  & 1                             & 0                             & 0                            \\
Hajek      &  & 3                        & 0                       & 1                      & 3                      &  & 1                             & 3                             & 2                            &  & 0                             & 0                             & 1                            \\
Kovanda    &  & 3                        & 0                       & 2                      & 2                      &  & 2                             & 2                             & 2                            &  & 0                             & 0                             & 1                            \\
Jires      &  & 2                        & 0                       & 2                      & 1                      &  & 2                             & 1                             & 1                            &  & 0                             & 0                             & 1                            \\ \bottomrule
\multicolumn{14}{l}{}\\[-0.75\normalbaselineskip]%
\multicolumn{14}{l}{Pre = pre-existing, RT = resource transfer, Co = collaboration.}%
\end{tabular}%
}
\end{table}

\subsection{Some mechanistic hypothesizing}
The network analyzes are viewed as qualitative explorations based on quantitized notions of relationships between actors.
These explorations may be used to formulate mechanistic hypotheses regarding the workings and nature of corruption structures.
On the basis of the foregoing, the following conjectures are made:
\begin{enumerate}
  \item Pre-existing close-knit modules among potentially corrupt officials form a sufficient basis for spurring a corruption network;
  \item Resource transfers within a corruption network will tend to a bipartite structure between officials and non-officials;
  \item Lovers or members with strong familial bonds that are part of the same corruption network act as a single operational unit within that network.
\end{enumerate}
Further research may shed light on these conjectures.

\section{Discussion}
\label{SEC:DISCUSS}
This study was motivated by the observation that studies of corruption often use network terminology but rarely use tools stemming from network science.
This small contribution then focused on studying corruption networks from the standpoint of multiplex (social) network analysis.
It offers a perspective on shared and differential adjacency matrices between pairs of layers in a multilayered network structure.
These matrices then serve as the basis for insightful visualization and the simple extension of a classic network centrality metric to the case of multilayered networks.
Using these amenities we provided an analysis of the Czech Rath affair that may be considered an addition to the analysis by Divi\'{a}k \emph{et al.} \cite{DDS18}.

This study deals with several limitations.
First, as also indicated by Divi\'{a}k \cite{D2018}, we are dealing with essentially covert networks.
Hence, we can never be sure that all relevant actors and ties have been identified.
This problem is analogous to the problem of identifying dark numbers in corruption measurement in general \cite{HLPmeas,RRpeeters}.
Second, the metrics developed here highlight only one aspect of network topology and are limited to pairings of layers in multilayered network structures.
Two natural inroads for further research could then be the extensions to alternative network metrics and more than two layers.
Moreover, we have considered only unweighted, undirected networks in the present work.
Another research area would be the study of multiplexity metrics in weighted and/or directed networks.

Notwithstanding these limitations, network analysis is a fruitful approach to studying corruption from a mechanistic perspective.
It might be timely indeed, when so much of the corruption literature invokes its nomenclature, to start using the tools of network science more widely.
\emph{This is a call to arms.}

\appendix
\section{Differential and shared adjacency matrices}
\label{APP:AdjacencyMatrices}
Consider the basic adjacency matrix given in (\ref{Eq: DiffAdjExp1}).
The basis of our differential metrics is then the difference in relation-specific adjacency matrices:
\begin{definition}[Differential Adjacency Matrix]\label{Def: DiffAdj}
    Let $\mathbf{A}_r$ and $\mathbf{A}_{r'}$ represent relation-specific adjacency matrices for relationship-types $r$ and $r'$, respectively.
    The \emph{differential adjacency matrix} between relationship-types $r$ and $r'$ is then the signed adjacency matrix
    \begin{equation}\label{Eq: DiffAdj}\nonumber
    \mathbf{A}_{rr'}^{-} = \mathbf{A}_r - \mathbf{A}_{r'},
    \end{equation}
    such that
     \begin{equation}\label{Eq: DiffAdjExp}\nonumber
    (\mathbf{A}_{rr'}^{-})_{ij} = \begin{cases}
                                 1 & \mbox{if} ~Y_{i} - Y_{j} ~\mbox{in relation-type} ~r \wedge Y_{i} \centernot{-} Y_{j} ~\mbox{in relation-type} ~r'\\
                                 -1 & \mbox{if} ~Y_{i} \centernot{-} Y_{j} ~\mbox{in relation-type} ~r \wedge Y_{i} - Y_{j} ~\mbox{in relation-type} ~r'\\
                                 0 & \mbox{otherwise}
                                \end{cases}.
    \end{equation}
\end{definition}

In Definition \ref{Def: DiffAdj} the $0$-indication represents the edges present or absent in both relationship-types.
For visualizing and quantifying the shared network, it is of interest to define the adjacency matrix consisting of edge-elements shared between classes:
\begin{definition}[Shared Adjacency Matrix]\label{Def: ShareAdj}
    The \emph{shared adjacency matrix} between relationship-types $r$ and $r'$ is the adjacency matrix $\mathbf{A}_{rr'}^{+}$ such that
     \begin{equation}\label{Eq: DiffAdjExp}\nonumber
    (\mathbf{A}_{rr'}^{+})_{ij} = \begin{cases}
                                 1 & \mbox{if} ~Y_{i} - Y_{j} ~\mbox{in relation-type} ~r \wedge Y_{i} - Y_{j} ~\mbox{in relation-type} ~r'\\
                                 0 & \mbox{otherwise}
                                \end{cases}.
    \end{equation}
\end{definition}
Note that both definitions may also be of use in subclass-specific networks in which the nodes represent random variables when we would think of the subscript $r$ as a group or subclass indicator \cite{FusedRidge,IMMEDIAD}.

\section{Software and data}
\label{APP:SoftData}
All analyzes were performed in \textsf{R} \cite{R} using the \texttt{rags2ridges} \cite{rags2ridges} and \texttt{igraph} \cite{igraph} packages.
All are freely available from the Comprehensive \textsf{R} Archive Network: \url{https://cran.r-project.org/}.
The data (in the form of adjacency matrices) as well as the coding script are available from the following GitHub page: \url{https://github.com/CFWP/LAL}.



\section*{Klip en klaar: An afterword}
In keeping with the tradition of making Liber Amicorum contributions a hotchpotch of recollection, memoir and science I will close with a personal afterword.
One that bears as the heading a rather archaic Dutch saying: \emph{klip en klaar}.
This saying is used, much to the amusement of others, quite often by Leo.
It means so much as to say that a given situation is clear beyond crystal and, when pronounced, draws images of the Low Countries in an era gone by.
Why, then, would I use such a phrase?
I hope to make this clear in what follows.

I have many anecdotes about Leo.
I could recount that he hired me as an assistant on the basis of a (poorly, ask my wife) handwritten CV because Karin Lasthuizen told him that this would be a good idea.
I could recount our musings regarding the eastern part of the Netherlands (where we are both from and which, by the way, is just as flat as any other part).
Instead, I am going to tell you something that I remember very vividly, something very telling about Leo.

It was a Friday night in 2005 in an empty office housing the Department of Public Administration.
I had worked in the department for years as a student and research assistant.
I was printing out my final MSc thesis in Political Science and Public Administration and was about to leave to go immerse myself in mathematics and statistics in Leuven (Belgium, less flat).
I remember Leo walking up to me, knowing that I was leaving, and giving me this great hug while wishing me all the best (even though he has some form of math-phobia; sorry about the above).
He gave me the best example of inspiration: kindness and belief in someone.

So there we have it.
I am a statistician and psychometrician working in the field of high-dimensional statistics and statistical machine learning.
And in this role I am very much indebted to Leo the public administrationist.
Both in the way I approach my science and in my choice of some of the subjects that I wish to explore.
As the very best scholars indeed do, he has sparked inspiration that transcends the boundaries between fields of study.
He has shown me by example how to be a scientist over a careerist and in many ways was my scientific father.
I can only hope to be such an example for my own scientific offspring.
\emph{Klip en klaar}.
\vspace{1cm}
\addresseshere


\begin{thebibliography}{10}
\bibitem{Battiston} Battiston, F., Nicosia, V., \& Latora, V. (2014).
    Structural measures for multiplex networks.
    \emph{Physical Review E}, 89: 032804.

\bibitem{FusedRidge} Bilgrau, A.E., Peeters, C.F.W., Eriksen, P.S., B{\o}gsted, M., \& van Wieringen, W.N/ (2015).
    Targeted fused ridge estimation of inverse covariance matrices from multiple high-dimensional data classes.
    arXiv:1509.07982 [stat.ME].
    \url{https://arxiv.org/abs/1509.07982}.

\bibitem{Bocca14} Boccaletti, S., Bianconi, G., Criado, R., del Genio, C.I., G\'{o}mez-Garde\~{n}es, J., Romance, M., Sendi\~{n}a-Nadal, \& Wang, Z. (2014).
    The structure and dynamics of multilayer networks.
    \emph{Physics Reports}, 544: 1--122.

\bibitem{BorgattiASN} Borgatti, S.P., Everett, M.G., \& Johnson, J.C. (2013).
    \emph{Analyzing Social Networks}.
    London: Sage.

\bibitem{C18} Chang, Z. (2018).
    Understanding the corruption networks revealed in the current Chinese anti-corruption campaign: A social network approach.
    \emph{Journal of Contemporary China}, 27: 735--747.

\bibitem{igraph} Cs\'{a}rdi, G. \& Nepusz, T. (2006).
    The igraph software package for complex network research.
    \emph{InterJournal, Complex Systems}, 1695.

\bibitem{RankingNC} De Domenico, M., Sol\'{e}-Ribalta, A., Omodei, E., G\'{o}mez, S., \& Arenas, A. (2015).
    Ranking in interconnected multilayer networks reveals versatile nodes.
    \emph{Nature Communications}, 6: 6868.

\bibitem{GHS18} de Graaf, G., Huberts, L.W.J.C., \& Str\"{u}wer, T. (2018).
    Integrity violations and corruption in Western public governance: Empirical evidence and reflection from the Netherlands.
    \emph{Public Integrity}, 20: 131--149.

\bibitem{IMMEDIAD} de Leeuw, F.A., Peeters, C.F.W., Kester, M.I., Harms, A.C., Struys, E.A., Hankemeier, T., van Vlijmen, H.W.T., van der Lee, S.J., van Duijn, C.M., Scheltens, P., Demirkan, A., van de Wiel, M.A., van der Flier, W.M., \& Teunissen, C.E. (2017).
    Blood-based metabolic signatures in Alzheimer's Disease.
    \emph{Alzheimer's \& Dementia: Diagnosis, Assessment \& Disease Monitoring}, 8: 196--207.

\bibitem{DV99} Della Porta, D., \& Vanucci, A. (1999).
    \emph{Corrupt Exchanges: Actors, Resources and Mechanisms of Political Corruption.}
    New York: Aldine de Gruyter.

\bibitem{D2018} Divi\'{a}k, T. (2018).
    Sinister connections: How to analyse organised crime with social network analysis.
    \emph{AUC Philosophica et Historica}, 2: 115--135.

\bibitem{DDS18} Divi\'{a}k, T., Dijkstra, J.C., \& Snijders, T.A.B. (2018).
    Structure, multiplexity, and centrality in a corruption network: the Czech Rath affair.
    \emph{Trends in Organized Crime}. \url{https://doi.org/10.1007/s12117-018-9334-y}.

\bibitem{FRalgo} Fruchterman, T.M.J., \& Reingold, E.M. (1991).
    Graph Drawing by Force-Directed Placement.
    \emph{Software: Practice \& Experience}, 21: 1129--1164.

\bibitem{GNalgo} Girvan, M., \& Newman, M.E.J. (2002).
    Community structure in social and biological networks.
    \emph{Proceedings of the National Academy of Sciences of the United States of America}, 99: 7821--7826.

\bibitem{PeopleOverviewGolis} Golis, O. (22/3/2018).
    \v{S}est let kauzy Rath j\'{\i} zdevastovalo \v{z}ivot, to je dostatecn\'{y} trest, rekla advok\'{a}tka administr\'{a}torky Novansk\'{e}
    [Six years of the Rath Case devastated her life, which is sufficient punishment, Novanska's lawyer said].
    \emph{iROZHLAS},
    \url{http://irozhl.as/2jv}.

\bibitem{Hill10} Hiller, P. (2010).
    `Understanding corruption: How systems theory can help',
    in: G. de Graaf, P. von Maravic, \& P. Wagenaar (Eds.).
    \emph{The Good Cause: Theoretical Perspectives on Corruption},
    Opladen: B. Budrich, pp. 64--82.

\bibitem{Leo98} Huberts, L.W.J.C. (1998).
    What can be done against public corruption and fraud: Expert views on strategies to protect public integrity.
    \emph{Crime, Law \& Social Change}, 29: 202--224.

\bibitem{Leo2000} Huberts, L.W.J.C. (2000).
    Anticorruption strategies: The Hong Kong model in international context.
    \emph{Public Integrity}, 2: 211--228.

\bibitem{H14} Huberts, L. (2014).
    \emph{The Integrity of Governance: What It Is, What We Know, What Is Done, and Where to Go}.
    New York: Palgrave Macmillan.

\bibitem{Leo18} Huberts, L.W.J.C. (2018).
    Integrity: What it is and why it is important.
    \emph{Public Integrity}, 20 (Sup 1): S18--S32.

\bibitem{HLPmeas} Huberts, L., Lasthuizen, K., \& Peeters, C. (2006).
    `Measuring Corruption: Exploring the Iceberg',
    in: C. Sampford, A. Shacklock, C. Connors, F. Galtung (Eds.).
    \emph{Measuring Corruption},
    Hampshire [etc.]: Ashgate Publishing, pp. 265--293.

\bibitem{LPS99} Huberts, L.W.J.C., Pijl, D., \& Steen, A. (1999).
    `Integrity and Corruption',
    in: C. Fijnaut, E. Muller and U. Rosenthal (Eds.).
    \emph{Police: Studies on the Organization and its Functioning},
    Samsom, Alphen aan den Rijn, pp. 57--79.

\bibitem{LKY11} Lauchs, M., Keast, R. L. \& Yousefpour, N. (2011).
    Corrupt police networks: uncovering hidden relationship patterns, functions and roles.
    \emph{Policing and Society: An International Journal of Research and Policy}, 21: 110--127.

\bibitem{NewmanBook} Newman, M. (2010).
    \emph{Networks}.
    Oxford: Oxford University Press.

\bibitem{Newton} Newton Media Archive.
    \url{https://www.newtonmedia.cz/en/services/archive}.

\bibitem{NL15} Nicosia, V., \& Latora, V. (2015).
    Measuring and modeling correlations in multiplex networks.
    \emph{Physical Review E}, 92: 032805.

\bibitem{Osifo18} Osifo, O.C. (2018).
    A study of distorted network: A narrative literature review analysis.
    University of Vaasa Reports $\sharp$8.

\bibitem{rags2ridges} Peeters, C.F.W., Bilgrau, A.E., \& van Wieringen, W.N. (2019).
    \emph{\texttt{rags2ridges}: Ridge Estimation of Precision Matrices from High-Dimensional Data}.
    URL \url{https://CRAN.R-project.org/package=rags2ridges}. \textsf{R} package version 2.2.1.

\bibitem{RRpeeters} Peeters, C.F.W., G.J.L.M. Lensvelt-Mulders \& K. Lasthuizen (2010).
    A note on a simple and practical randomized response framework for eliciting sensitive dichotomous \& quantitative information.
    \emph{Sociological Methods \& Research}, 39: 283--296.

\bibitem{R} \textsf{R} Core Team (2014).
    \textsf{R}: A language and environment for statistical computing.
    \textsf{R} Foundation for Statistical Computing, Vienna, Austria.
    URL \url{https://www.R-project.org/}.

\bibitem{RibETAL18} Ribeiro, H.V., Alves, L.G.A., Martins, A.F., Lenzi, E.K., \& Perc, M. (2018).
    The dynamical structure of political corruption networks.
    \emph{Journal of Complex Networks}, 6: 989--1003.

\bibitem{Slinger18} Slingerland, W. (2018).
    \emph{Network Corruption: When Social Capital Becomes Corrupted}.
    PhD thesis: VU University Amsterdam.
\end{thebibliography}
\end{document}